%% file: paper.tex
\documentclass[10pt, conference]{IEEEtran}
\IEEEoverridecommandlockouts
\usepackage{cite}
\usepackage{amsmath,amssymb,amsfonts}
\usepackage{algorithmic}
\usepackage{graphicx}
\usepackage{subcaption}
\usepackage{textcomp}
\usepackage{xcolor}
\usepackage[font=small, skip=4pt]{caption}
\usepackage{tikz}

\def\BibTeX{{\rm B\kern-.05em{\sc i\kern-.025em b}\kern-.08em
    T\kern-.1667em\lower.7ex\hbox{E}\kern-.125emX}}
\begin{document}
\newcommand{\Sim}{\texttt{RDZsim}} 
\newcommand{\Tool}{\texttt{FlexRDZ}} 

\graphicspath{{./figs/}}

\newcommand\copyrighttext{%
\footnotesize \textcopyright 2025 IEEE. Personal use of this material is permitted.
Permission from IEEE must be obtained for all other uses, in any current or future
media, including reprinting/republishing this material for advertising or promotional
purposes, creating new collective works, for resale or redistribution to servers or
lists, or reuse of any copyrighted component of this work in other works.}

\newcommand\copyrightnotice{%
\begin{tikzpicture}[remember picture,overlay]
\node[anchor=south,yshift=10pt] at (current page.south)
{\fbox{\parbox{\dimexpr\textwidth-\fboxsep-\fboxrule\relax}{\copyrighttext}}};
\end{tikzpicture}%
}

\title{FlexRDZ: Autonomous Mobility Management for Radio Dynamic Zones\\
\thanks{This material is based upon work supported by the National Science Foundation under Grant Number~CNS-1827940. This work has been submitted to the IEEE for possible publication. Copyright may be transferred without notice, after which this version may no longer be accessible.}
} 

\author{\IEEEauthorblockN{Aashish Gottipati}
	\IEEEauthorblockA{\textit{University of Texas Austin}\\
		agottipati@utexas.edu}
	\and
	\IEEEauthorblockN{Jacobus Van der Merwe}
	\IEEEauthorblockA{\textit{University of Utah}\\
		kobus@cs.utah.edu}
}

\maketitle
\copyrightnotice

\input{abstract}
\begin{IEEEkeywords}
Radio Dynamic Zone, Mobility Management, Network Control, AI Planning
\end{IEEEkeywords}
\input{introduction}
\input{background}
\input{related}
\input{design}
\input{evaluation}
\input{conclusion}

\bibliographystyle{IEEEtran}
\bibliography{bibliography}

\end{document}

%% file: abstract.tex


\begin{abstract}
\Tool{} is an online, autonomous manager for radio dynamic zones (RDZ) that seeks to enable the safe operation of RDZs through real-time control of deployed test transmitters. \Tool{} leverages Hierarchical Task Networks and digital twin modeling to plan and resolve RDZ violations in near real-time. We prototype \Tool{} with GTPyhop and the Terrain Integrated Rough Earth Model (TIREM). We deploy and evaluate \Tool{} within a simulated version of the Salt Lake City POWDER testbed, a potential urban RDZ environment. Our simulations show that \Tool{} enables up to a 20 dBm reduction in mobile interference and a significant reduction in the total power of leaked transmissions while preserving the overall communication capabilities and uptime of test transmitters. To our knowledge, \Tool{} is the first autonomous system for RDZ management.
\end{abstract}



%% file: introduction.tex


\section{Introduction}
\label{sec:intro}

Radio technology has progressed tremendously in recent years, as seen by the proliferation of software-defined radios, 5G networks, and the advent of terra hertz technology. To continue to drive innovation, researchers require access to adequate radio test facilities, enabling them to develop, benchmark, and validate their test transmitters without worrying about potential impacts on nearby radio infrastructure. In accordance, the notion of developing a universal radio test facility led to the proposed concept of a National Radio Dynamic Zone (NRDZ)~\cite{nrdz-article}. Radio Dynamic Zones (RDZ) are isolated from the outside world spectrally, enabling the deployment of new test transmitters. These zones prevent internal transmissions from escaping, freeing test operators from the responsibility of worrying about interfering with nearby infrastructure. 

While an RDZ seeks to coexist with existing infrastructure, an RDZ also seeks to enable users to deploy and test new transmitter technology. In many cases, these test transmitters may be inimical towards nearby infrastructure, motivating the RDZ operator to employ oversight over test transmitters. While operators may be able to exercise oversight over stationary test transmitters due to their static position, mobile transmitters further tax an RDZ operator's ability to supervise the RDZ properly. For example, many spectrum allocation techniques assume that test transmitters are stationary, failing to model mobile entities accurately~\cite{win-eval, win-eval2}; however, in an RDZ, test transmitters may or may not be stationary (e.g.,\ naval and air radar systems). In these cases, utilizing traditional techniques may not be sufficient to prevent test transmitters from impacting nearby infrastructure. We, therefore, choose to explore RDZ mobility based on the increase in management complexity and the assumption that many future test transmitters will be mobile~\cite{drone-swarm}. 

\begin{figure}[tbp!]
    \centering
    {\includegraphics[width=0.8\columnwidth]{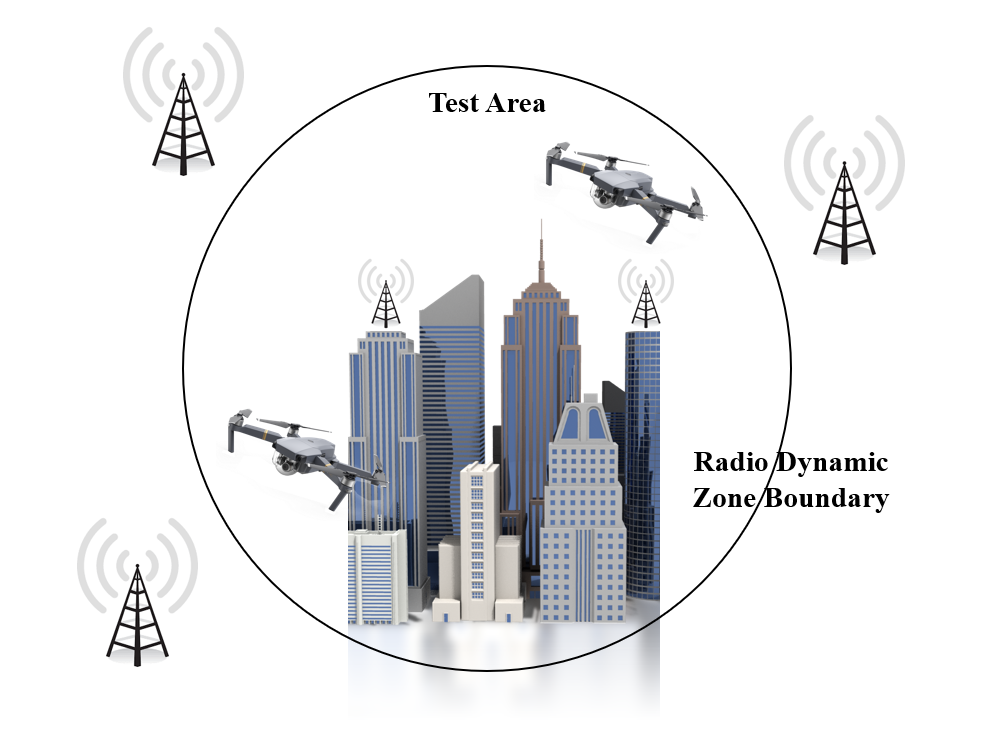}}
    \caption{Mobile RDZ Deployment.}
    \vspace{-0.7cm}
    \label{fig:s3}
\end{figure}

To simplify our model of transmitter mobility, like others~\cite{rdz-zme}, we assume a complete control framework, meaning that all transmitters cede control to a management entity, enabling a higher level of flexibility within the RDZ (see Figure~\ref{fig:s3}). For example, the area in which a mobile test transmitter can operate becomes more flexible through techniques such as spectrum sharing and real-time transmit power adjustments, since leakage and interference can be handled dynamically.

To realize a more flexible RDZ, we propose \Tool{}. \Tool{} is a closed-loop, autonomous RDZ manager that seeks to enable the safe operation of an RDZ through real-time control of deployed test transmitters. \Tool{} is primarily defined by its Hierarchical Task Network (HTN), a graph-based planning approach, and its RDZ digital twin. These two components enable \Tool{} to dynamically model the RDZ environment and generate a plan to maintain the ``health'' of the RDZ in the event a compromise occurs, e.g.,\ internal transmissions are detected beyond the RDZ boundary. \Tool{} utilizes its programmable control framework to execute its generated plan in order to preserve the health of the RDZ. To validate our implementation of \Tool{}, we deploy and evaluate \Tool{} within a simulated version of the Salt Lake City POWDER testbed~\cite{powder}, a potential urban RDZ environment. Our simulations show that \Tool{} enables up to a $20$ dBm reduction in mobile interference and a significant reduction in the total power of leaked transmissions while preserving the overall communication capabilities and uptime of test transmitters.

To realize an autonomous RDZ management system, we make the following contributions: 
\begin{trivlist}
	\vspace{-0.2cm}
	\item $\bullet$ The design of \Tool{}, a near real-time, autonomous RDZ manager that aims to maintain an RDZ in an online fashion.
	\item $\bullet$ A prototype implementation of \Tool{} that reduces RDZ leakage and preserves transmitter communication.
	\item $\bullet$ Quantitative results that demonstrate the benefits of \Tool{} by highlighting its performance and use cases.
	\vspace{-0.2cm}
\end{trivlist}

The remainder of this paper is laid out as follows. Section~\ref{sec:background} discusses key background information related to planning.  Section~\ref{sec:related} surveys related work in the space of dynamic spectrum environments. Section~\ref{sec:Design} discusses the design and implementation decisions pertaining to \Tool{}. Section~\ref{sec:evaluation} highlights \Tool{}'s capability within the context of autonomous RDZ management. Lastly, section~\ref{sec:conclusion} comments on the limitations of our approach, proposes future areas of research, and provides concluding remarks.



%% file: background.tex


\section{Background}
\label{sec:background}
AI planning refers to the problem of finding a set of actions that, if executed, will transition the environment from an initial state to a goal state. More formally, given a set of states $\mathcal{S}$ and actions $\mathcal{A}$, we require a mapping between states and actions $\pi : \mathcal{A} \times \mathcal{S} \rightarrow [0,1]$ where the actions push the agent in the direction of the goal state. Furthermore, the transition model defines the probability of a taking an action given the current state: $\pi(a, s) = Pr(a_t = a| s_t = s)$. We call the combination of the mapping and transition model the policy, i.e., a distribution of actions conditioned on the current environment state. This problem is canonically known as a Markov decision process (MDP).

Modern approaches to AI planning utilize a mix of deep reinforcement learning (RL), statistical techniques, and integer programming~\cite{dream-to-control} with the former becoming the dominant planning approach given its application in robot planning and manipulation~\cite{soft-arm}. At a high level, RL seeks to learn a policy for acting in an environment where the environment is represented by an MDP. In contrast to other deep learning disciplines, RL provides an agent with a reward function, which aims to guide the agent toward favorable states in the state space. Upon taking an action, the agent receives a reward according to a predefined reward function, which provides a metric for gauging the performance of an action. Long-term rewards are penalized via a discount rate, which prioritizes events in the immediate future over the distant future. In essence, the agent seeks to learn a policy by maximizing the returned reward, which, in turn, guides the agents toward favorable states. 

However, there are two issues with this formulation. One, while deep learning-based techniques perform well in practice, many of these techniques suffer from the lack of explainability~\cite{rl-explain}; hence, robust safeguards are required when deploying in environments with risky tail-end events. Two, in the real world, state observations are noisy, i.e.,\ states are not fully observable. In this case, the problem becomes a partially-observable MDPs (POMDP), a well-studied problem in robotics~\cite{visio-motor}, but requires more assumptions for safe deployments.

\begin{figure}[t!]
	\centerline{\includegraphics[width=0.9\columnwidth]{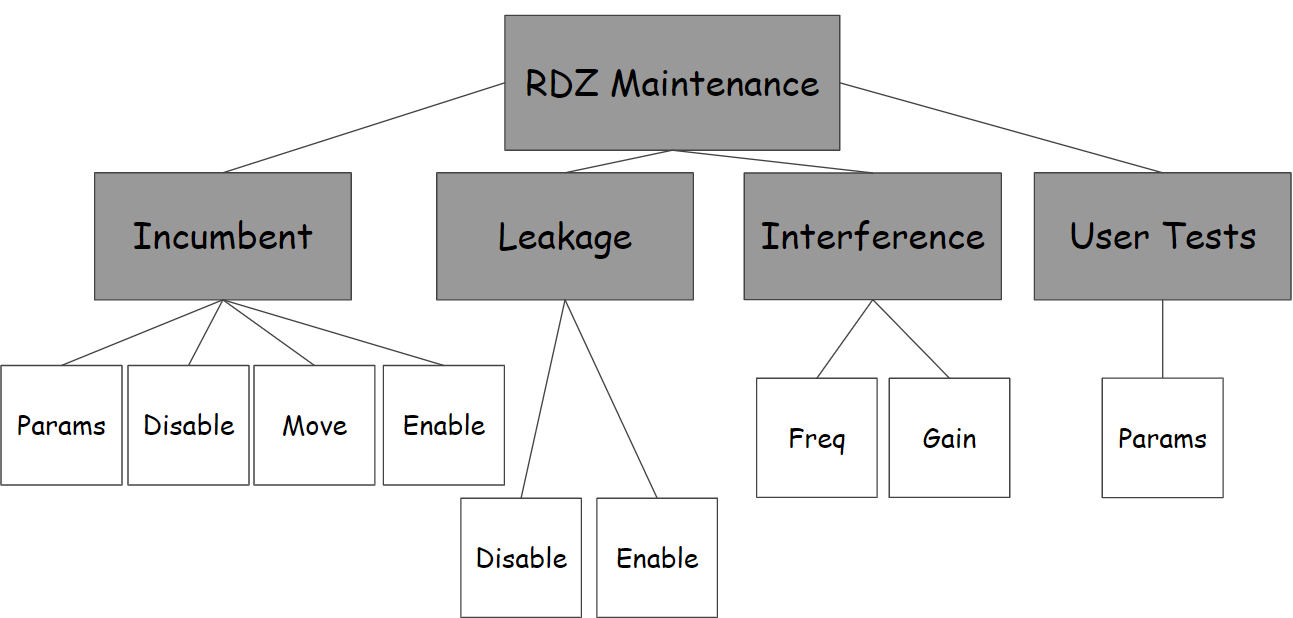}}
	\caption{\Tool{}'s AI Planner.\\RDZ maintenance is divided into four subtasks. For example, to satisfy an incumbent's request to adjust the area of the RDZ, the HTN returns the following atomic tasks: update the RDZ parameters (e.g.,\ boundary coordinates), disable non-compliant transmitters, move mobile transmitters to the new zone, and enable all compliant transmitters.}
	\vspace{-0.5cm}
	\label{fig:htn}
\end{figure}

In the context of RDZs, transparency and safety are crucial since an incorrect plan could broadly impact surrounding telecommunications services~\cite{rdz-zme}. Given this, we focus on a classical planning approach known as HTNs~\cite{htn-app}. HTNs are dependency graphs that are constructed with domain knowledge, meaning that the policy $\pi$ is hand-crafted rather than learned. In contrast to other modern methods, HTNs are human interpretable, offering more transparency into understanding why certain decisions or plans were generated~\cite{htn-transparent}. HTNs follow a hierarchical structure where the root node corresponds to the initial state or the start task. The starting task is then recursively broken down into subtasks until an atomic action can be executed (see Figure~\ref{fig:htn}). The root node connects to its descendants via edges, which represent dependencies. To transition from the initial state (root node) to the goal state, the root's dependencies must be satisfied by carrying out the corresponding atomic actions. When called upon, an HTN generates a sequence of atomic actions to satisfy an overarching goal. We seek to leverage HTNs to dynamically generate plans to mitigate RDZ violations and autonomously maintain the RDZ environment.

%% file: related.tex


\section{Related Work}
\label{sec:related}


\begin{figure}[tbp!]
	\centerline{\includegraphics[width=0.8\columnwidth]{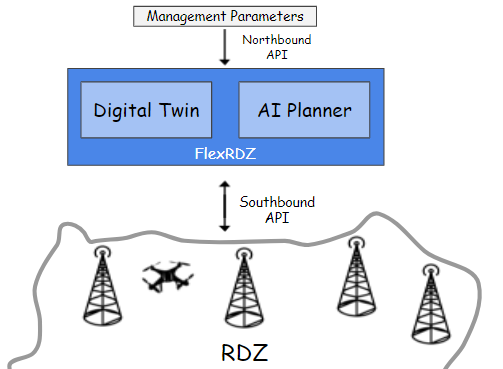}}
	\caption{\Tool{} Architecture.}
	\vspace{-0.8cm}
	\label{fig:optool}
\end{figure}

\subsection{Citizens Broadband Radio Service}
{\bf Interference Management.} To mitigate interference within CBRS, the Wireless Innovation Forum (WINNFORM) suggest three different graph-based approaches to spectrum allocation~\cite{winn-1, winn-2, winn-3}. 
Approaches one and two utilize a graph coloring algorithm to allocate non-overlapping spectrum to nearby transmitters, while approach three employs a recursive clustering algorithm. Gao et al.\ evaluate all three approaches using a suite of propagation models and GIS map data of Virginia Beach and San Diego~\cite{win-eval,win-eval2}, showing that all three approaches are indeed effective at reducing potential interference. In contrast to WINNFORM, Abbass et al. explore the application of Q-learning for spectrum allocation in CBRS~\cite{cbrs-q}. Specifically, the authors investigate opening up idle access priority access license (PAL) channels to general authorized access (GAA) users. Numerical results demonstrate improvements in spectrum utilization and data rate per unit cost; however, real-world evaluations are necessary.  

Although, while the WINNFORM approaches excel at mitigating interference, there are many drawbacks in practice. For example, since these approaches utilize the overlap in deployment area as an estimate for interference between transmitters, the estimates of interference may underestimate or overestimate the actual interference based on the deployed environment~\cite{under-over-graph}. In addition, graph coloring is an NP-complete problem, and, as such, most algorithms are subject to non-polynomial growth, meaning that, with dense transmitter deployments, the computation layer becomes the bottleneck~\cite{under-over-graph}. Lastly, these interference management techniques yield their highest performance when transmit nodes are stationary, which may not hold in an RDZ environment where test transmitters may be mobile, e.g.,\ drones. Thus, to better model dynamic RDZ environments, we supplement our approximations with simulations, leading to the addition of \Tool{}'s digital twin. We note that we are not solving the interference management problem within CBRS; instead, we build upon techniques used in CBRS and generalize them to RDZs.

{\bf Mobility Management.} The most relevant work to handling mobility within an RDZ-like environment is related to detecting naval incumbents in CBRS. To detect naval incumbents, the National Telecommunications and Information Administration (NTIA) proposed the use of an environmental sensing capability (ESC) network~\cite{ntia}. An ESC network comprises multiple sensors employed to detect an incumbent's presence and trigger protective measures upon detection. Nguyen et al. formulated the ESC deployment problem as a set cover problem to compute the minimum number of sensors to cover an area of interest while minimizing the overlapping area between sensors, since overlapping areas may lead to false positives~\cite{mobi-esc}. As opposed to preemptively turning off nearby equipment upon detecting an incumbent, Kang et al.\ offer a different approach: using a management entity to oversee interactions between transmitters~\cite{lte-tdd}. Instead of proactively pausing communications in inflection areas (e.g., where naval incumbents are detected), Kang suggests employing more dynamic techniques such as spectrum sharing and virtualization. In our formulation of an RDZ, we implicitly cover both of these cases; however, we emphasize that we are not bound by the same assumptions as CBRS.

\subsection{Radio Dynamic Zones}
We now cover related work on RDZs. Maeng et al. propose a spectrum monitoring approach for out-of-zone signal leakage detection and explore spatial correlation-based estimation techniques for signal prediction~\cite{rdz-ooz}. The authors constrain their RDZ formulation geographically and assume sensor nodes that cover and monitor the boundary of the RDZ. Through simulation, the authors demonstrate that their spatial correlation-based algorithm enables a larger RDZ radius with sparsely deployed sensor nodes in comparison to propagation-loss techniques. In addition, Maeng et al. present an RDZ concept that relies on both autonomous aerial and ground sensor nodes for radio environment monitoring, enabling real-time radio environment maps of all relevant frequencies and locations~\cite{rdz-rem}. Lastly,~\cite{rdz-zme} discusses key challenges for real RDZ deployments and details a Zone Management Engine for RDZ supervision. Specifically, the Zone Management Engine is composed of a decision engine and three subsystems: spectrum, experiment, and policy management systems. In concert, these three subsystems provide the decision engine with experiment information, spectrum policy rules, user information, and resource allocations to enable dynamic real-time coordination of systems within the zone coupled with user interference protection. In other words, \Tool{} is a realized prototype of the previously proposed Zone Management Engine~\cite{rdz-zme}.


We diverge from previous work in fundamental ways. First, we primarily focus on the problem of streamlining RDZ maintenance through dynamic, flexible planning practices rather than spectrum monitoring. Second, we assume infrastructure for spectrum monitoring and leverage real-time environment maps to estimate and plan for future transmitter behavior. Lastly, we approach RDZ maintenance through a systems perspective, prototyping and evaluating a system for real-time RDZ maintenance and control.



%% file: design.tex


\section{Design and Prototype}
\label{sec:Design}


\subsection{Overview}

\Tool{} is an operational tool that seeks to mitigate RDZ violations through swift resolution and maintain the environment in an autonomous and online fashion. RDZ violations encompass situations that compromise the state of the RDZ (e.g.,\ transmissions are detected beyond the environment boundary or the RDZ fails to accommodate local infrastructure) and cases where user test objectives are conceded. 

As seen in Figure~\ref{fig:optool}, \Tool{} receives management parameters via its northbound interface. These management parameters encompass RDZ maintenance variables (e.g.,\ boundary parameters, leakage thresholds, and interference thresholds) and user test objectives (e.g.,\ reserved areas and frequencies). Upon failing to maintain these parameters, \Tool{} utilizes its planner to generate a plan and resolve the situation dynamically. 

Lastly, \Tool{} functions in an online, autonomous fashion. These characteristics are a byproduct of \Tool{}'s southbound interface, which routinely pulls the state of the RDZ from the environment and enables direct control over test transmitters. Illustrated in~\cite{rdz-rem}, real-time RDZ state can be monitored via ground-based and aerial sensing infrastructure. By retrieving real-time updates, \Tool{} can, in near real-time, generate solutions and resolve environmental issues with its planner. The combination of \Tool{}'s southbound and northbound interface enables it to function within a closed-loop, online manner. The architecture of \Tool{} is in-line with previously envisioned RDZ supervisors, e.g., a Zone Management Engine~\cite{rdz-zme}. 

\subsection{Intelligent Control}

While planning may deceptively appear straightforward, it is important to note that RDZ environments quickly become too convoluted to manage due to the sheer number of observable environment states over time, making it infeasible to iterate over all potential control solutions. 
Although this problem is challenging, a great deal of related work has sought to overcome similar complexities in adjacent problems by efficiently searching over the planning space through the use of symbolic methods, artificial intelligence (AI), and various optimization techniques~\cite{dream-to-control}. Like other planning techniques, symbolic methods such as HTNs are utilized to search over the planning space efficiently~\cite{htn-app}. We leverage HTNs for their increased decision transparency and reliability-- two characteristics essential for maintaining an urban RDZ. We note that the design of \Tool{} does not preclude the use of other planning techniques.

\subsection{Digital Twin Modeling}

In combination with GIS map data and transmitter parameters, \Tool{} leverages an internal radio-frequency (RF) model to ``digitize'' the RDZ environment and generate a radio-environment map in real-time. \Tool{}'s RF model serves to model the RF interactions among transmitters, estimate the coverage for a transmitter, and derive RDZ-specific, key performance indicators. Note that the design of \Tool{} does not rely on any one modeling technique and can generalize to various methods such as path-loss or AI-based. By encoding the environment, \Tool{} can estimate potential states of the RDZ by simulating future actions through its RF model, enabling more robust control updates.



\subsection{Implementation}
To prototype \Tool{}, we utilized the Terrain Integrated Rough Earth Model (TIREM) propagation model~\cite{tirem}. TIREM is a set of physics-based algorithms used to estimate the coverage for mobile land radios and point-to-point distances. We opted to utilize TIREM as it is the standard propagation model utilized by the United States government. However, through preliminary evaluations, we observed that modeling the entire RDZ environment via TIREM can be extremely costly. In our case, we observed estimation times of approximately $30$ seconds for large areas, e.g.,\ the downtown Salt Lake City area, preventing real-time control. Thus, similar to the idea proposed in~\cite{dl-prop}, to alleviate this bottleneck, we trained a neural network to approximate the RF maps produced by TIREM, cutting inference time to approximately $70$ ms, and enabling real-time control. We opted to utilize a fusion-based network for their ability to learn more intricate feature representations~\cite{multi-modal}.

As input, the RF model accepts GIS map data that corresponds to the RDZ deployment area and the transmit parameters of the mobile transmitter. The GIS map data is processed through a series of ResNet blocks~\cite{resnet}, while transmit parameters are encoded via fully-connected layers. The two embeddings are concatenated and used as input to another fully-connected layer to generate a fusion embedding. From this embedding, the model learns a decoding function to output the predicted RF map.

\begin{figure}[tbp!]
    \centering
    \begin{subfigure}[b]{0.5\columnwidth}
        \centering
        {\includegraphics[width=1.0\columnwidth]{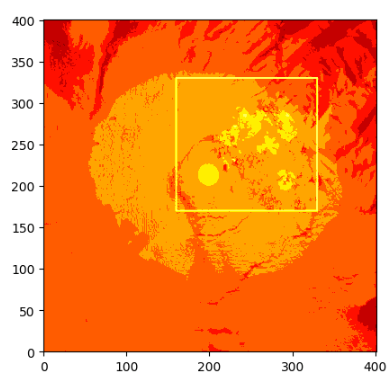}}
        \caption{}
        \label{fig:rssi-sim}
    \end{subfigure}%
    ~ 
    \begin{subfigure}[b]{0.5\columnwidth}
        \centering
        {\includegraphics[width=1.0\columnwidth]{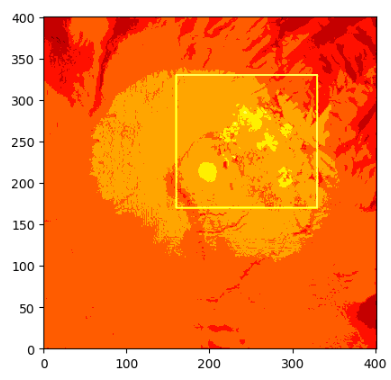}}
        \caption{}
        \label{fig:rssi-net}
    \end{subfigure}
    \caption{Simulated RF Map vs. Generated RF Map.\\The left figure represents the path-loss contour output of our simulation tool while the right figure represents the path-loss contour output of our model.}
    \label{fig:output-maps}
        \vspace{-0.7cm}
\end{figure}

To train our model, we first generated $10000$ training instances. A training instance consisted of the GIS map data, the mobile transmit parameters, and the ground truth TIREM RF map of the downtown Salt Lake City area. After generating the dataset, we randomly split the dataset into train, validation, and test set splits following an $80/10/10$ ratio. We empirically chose the fusion embedding size of $2048$ as it resulted in the fast training times and an overall storage footprint of approximately $0.8$ Mb. 
Furthermore, we trained our model for $150$ epochs via Adam with an initial learning rate of $0.001$, using mean-squared error (MSE) as our loss function, until we observed model convergence on our validation set. We verified that our model generates reasonable RF maps, producing an average empirical error of $-0.045$ dBm per cell (see Figure~\ref{fig:output-maps}). We emphasize that our learned model is unlikely to generalize to new settings as our training data is drawn from the Salt Lake City area. We leave learning a general model as an area for future work. 

Furthermore, to realize \Tool{}'s HTN planning component, we leveraged the open-source HTN planner known as GTPyhop~\cite{pyhop}. GTPyhop provides a basic framework for constructing HTNs, enabling users to specify objectives, sub-tasks, and action primitives. GTPyhop utilizes a modified depth-first search to search over the planning space efficiently. Nau et al. provide an in-depth analysis of the GTPyhop planning algorithm~\cite{pyhop}. Moreover, we constructed \Tool{}'s HTN planner (see Figure~\ref{fig:htn}) based on a small set of atomic primitives: idle, disable transmitter, enable transmitter, and round-robin frequency reassignment, which are routinely utilized to manage dynamic spectrum environments to foster generalizability. Like other works~\cite{digital-twin}, we intertwined the digital twin with the HTN to produce a more robust management policy~\cite{digital-twin}. For example, suppose that a transmitter has been disabled due to leaked signals. To re-enable the transmitter, \Tool{} must ensure that the newly enabled transmitter does not violate the terms of the RDZ. By leveraging its digital twin, \Tool{} can simulate the impact of its future decision; hence, the transmitter is only re-enabled when its digital counterpart does not leak, which leads to a more compliance-oriented policy.



%% file: evaluation.tex


\section{Evaluation}
\label{sec:evaluation}


\subsection{Setup}
We utilized our internal RDZ simulation tool to simulate an RDZ deployed within the POWDER testbed. We opted to simulate the POWDER testbed environment as POWDER may provide RDZ functionality to researchers in the future. In addition, the testbed is situated near the downtown Salt Lake City urban area, which directly reflects the urban RDZ deployment scenario laid out in Section~\ref{sec:intro}. 

The simulated environment consisted of the Salt Lake City area projected into a $400 \times 400$ matrix. Each entry in the projected matrix corresponded to the summation of predicted TIREM signal strengths of the deployed transmitters at that given point. Additionally, we leveraged GIS map data of the Salt Lake City area and approximations of the POWDER testbed's campus buildings to model the terrain of the POWDER testbed. The parameters of the digital twin are fixed; hence, we leave dynamic digital twin modeling-- utilizing real-time updates to adjust the parameters of the digital twin-- for future work. The evaluation seeks to evaluate the performance of \Tool{}'s AI planner (see Figure~\ref{fig:htn}) rather than the performance of \Tool{}'s digital twin. Note that we present a suite of planning-based techniques and seek to demonstrate the efficacy of \Tool{} to bolster RDZ integrity against non-management based environments, e.g., an environment with no planning agent.

To evaluate our HTN planner, we compare with the following planning approaches. 
\begin{enumerate}
    \item {\it Stochastic HTN } The stochastic HTN generates an identical plan to our HTN implementation; however, the agent executes an action with probability $1 - \epsilon$. We set $\epsilon$ to $0.1$, $0.2$, $0.3$ ($1$, $2$, and $3$ respectively). 
    \item {\it Proximal Policy Optimization (PPO).} A state-of-the-art model-free RL algorithm. 
    \item {\it Random.} The agent adopts a random policy and executes an action at random.
    \item {\it Naive.} No planning agent. In this scenario, the RDZ operator adopts a trust-based policy (users will seek to not violate the terms of the RDZ); hence, no control framework is utilized. This is the primary baseline for comparison.
\end{enumerate}


\begin{figure*}[tbp!]
    \centering
    \begin{subfigure}[t]{0.5\textwidth}
        \centering
        {\includegraphics[width=1.0\columnwidth]{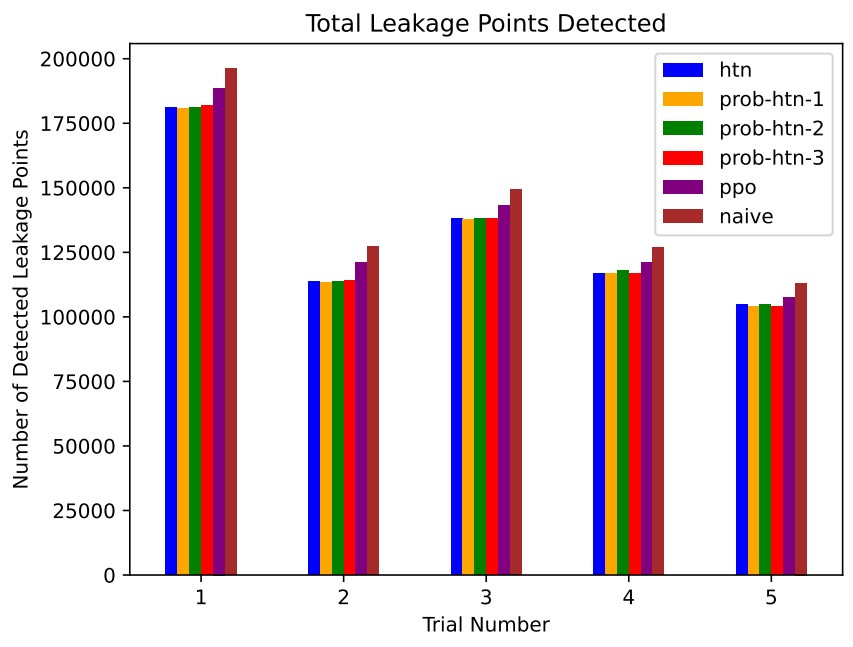}}
        \caption{}
        \label{fig:leak_points}
    \end{subfigure}%
    ~ 
    \begin{subfigure}[t]{0.5\textwidth}
        \centering
        {\includegraphics[width=1.0\columnwidth]{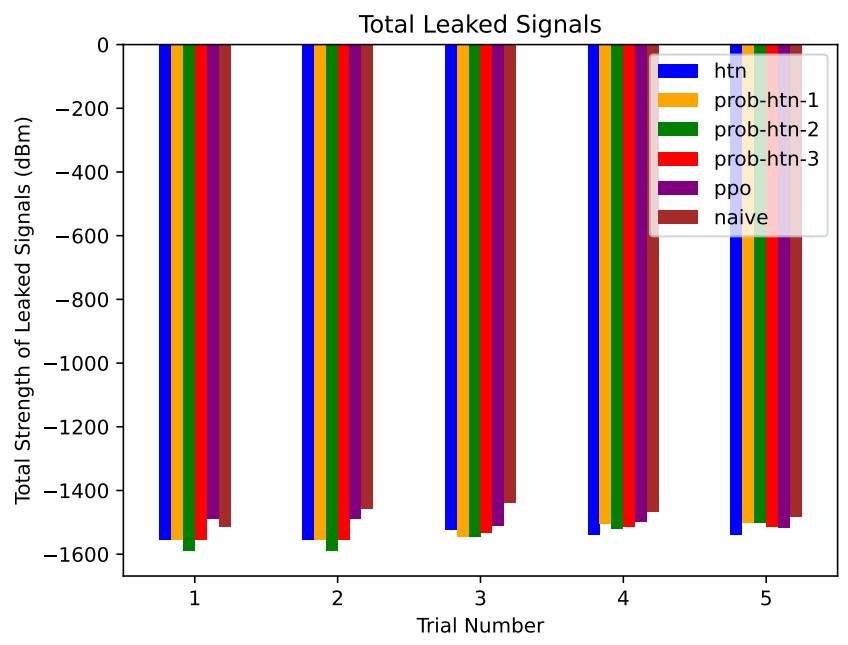}}
        \caption{}
        \label{fig:leakage_signal}
    \end{subfigure}
    \caption{Observed Mobile Leakage. A smaller number indicates less RDZ leakage, while a larger number indicates more leakage.}
    \label{fig:leakage}
\end{figure*}

Furthermore, the simulated RDZ was defined by a rigid geographical boundary, which specified where signals above a given power must not be detected, mimicking our desired urban RDZ deployment. The deployment consisted of $10$ fixed endpoints deployed at a height of $1.8$m with an antenna gain of $-2$ dBi, $9$ rooftop stationary test transmitters deployed at a height between $20$m and $40$m with an antenna gain of $4.9$ dBi, and $8$ densely deployed, stationary test transmitters placed at a height of $8$m with an antenna gain of $4.9$ dBi. The deployed transmitters were evenly split among the frequencies of $3600$, $3610$, and $3620$ MHz. The placement and parameters of transmitters were selected to mimic the current POWDER testbed deployment. A single mobile transmitter was deployed at a height of $30$m with an antenna gain of $4.9$ dBi, on one of the previously mentioned frequencies, depending on the simulated evaluation. We justify our limited evaluation with only one mobile node since in an early stage POWDER RDZ deployment, we expect the number of mobile transmitters to be constrained by the POWDER testbed deployment area, the density of existing stationary transmitters, and the restricted set of operating parameters.

To measure the performance of our system, we track the following metrics across trials. 
\begin{enumerate}
    \item {\it Leakage Points.} The number of locations outside the RDZ boundary with a value above the designated RDZ power threshold.
    \item {\it Total Strength of Leaked Signals.} The total power of induced signals outside the RDZ boundary with a value above the designated RDZ leakage threshold.
    \item {\it Induced Interference.} The total mobile interference observed within the RDZ boundary for a given trial.
    \item {\it Mobile Signal-to-Interference-plus-Noise Ratio.} The observed SINR during each evaluation step.
    \item {\it Mobile Transmitter Uptime.} The percentage of valid time steps in which the transmitter is active. A time step is considered valid if the mobile transmitter is not violating the policies defined by the RDZ operator (e.g.,\ mobile transmissions are limited to the area of the RDZ).
\end{enumerate}

\subsection{RL Training Procedure}
To benchmark our HTN approach against PPO, we leverage Open AI Stable Baselines and Open AI Gym to train an agent for RDZ maintenance. We adopt the following RL formulation.

{\bf State.} A state observation corresponds to a $2048$ dimensional fusion vector that encodes information about the current state of the simulated RDZ environment as well as the mobile transmit parameters. We leverage the fusion backbone of our RF model to encode the state of the simulation.

{\bf Action.} A discrete value between $0$ and $3$. We map each action of our HTN to a number between $0$ and $3$; specifically, $0$ corresponds to idle, $1$ disable transmitter, $2$ enable transmitter, and $3$ round robin frequency assignment.

{\bf Reward.} Our reward is defined per time step as 
\begin{equation} 
\label{eq:reward-fn}
10 \times U + \frac{S}{30} - \frac{I}{I_T} - \frac{P}{A} - \frac{L}{L_T}
\end{equation}

where $U$ corresponds to the number of time steps where the mobile transmitter is enabled, $S$ is the SINR of the mobile transmitter in dB, $I$ is the induced mobile interference in dBM, $I_T$ is the interference threshold in dBM, $P$ is the number of leakage points detected, $A$ is the total area of the RDZ in units, $L$ is the total power of leaked signals in dBM, and $L_T$ is RDZ power threshold for leaked signals in dBM. While simple, our reward function encourages uptime and open communication channels, while minimizing mobile interference and leakage, all of which are critical to maintaining an RDZ. Note that we clip each reward component by an empirically chosen constant to better shape the reward function and reduce asymmetries between goals.

We represent the learned policy with a multi-layer perceptron (MLP). We trained our agent in simulation for $8000$ episodes via PPO~\cite{ppo}, a state-of-the-art on-policy training approach, with each episode consisting of $10$ time steps. After each time step, we apply the agent's action where the resulting reward is calculated according to equation~\ref{eq:reward-fn} until policy convergence is observed. 

\subsection{Mobile Leakage}
For this evaluation, we set the RDZ leakage threshold to $-95$ dBm, indicating that if any signals were detected beyond the RDZ boundary above $-95$ dBm, then this constituted an RDZ violation, i.e.,\ leakage. Since our problem formulation considered an urban RDZ with nearby existing infrastructure, and signals with a strength of $-90$ dBm are likely to drown within the noise floor, signals at $-95$ dBm are likely to be too weak to impact nearby systems. 

The evaluation sought to evaluate \Tool{}'s HTN against other planning approaches in terms of leakage mitigation. The naive approach relies on ``good faith'' that the mobile transmitter will not violate the trust agreement of the RDZ. The evaluation procedure was as follows. A partition of POWDER nodes ($9$ nodes) was deployed on $3600$ MHz. A single mobile transmitter was then randomly deployed on $3600$ MHz. The mobile transmitter was then simulated for $50$ discrete time steps, with the transmitter moving $5$ units in a random direction at each time step. The number of leakage points detected and the total strength of leaked signals were recorded and summed at each step. We repeated the experiment for $5$ trials, reusing the same transmit and movement parameters for all planning methods.

We summarize the results of the evaluation in Figure~\ref{fig:leakage}. The results show that the HTN-based planners consistently led to less leakage in comparison to the RL-based and naive planning approaches. We omit the results from the random policy as the transmitter uptime was significantly lower than the other planning approaches (see Table~\ref{tab:uptime}). The decrease in leakage can be observed by the fact the first four bars in Figure~\ref{fig:leakage} are consistently lower than the last two, which correspond to PPO and the naive approach. It is important to note that the stochastic planners outperform \Tool{}'s deterministic planner in some cases, e.g.,\ trials $4$ and $5$ in terms of leakage points and trials $1$-$3$ in terms of leaked signals. This is likely due to the exploration vs. exploitation trade-off commonly discussed in RL literature~\cite{ppo}. In this case, due to the $\epsilon$ hyperparameter, the stochastic planners are able to travel to new states that are not reachable by the deterministic HTN. For example, since the deterministic HTN is reactive and not preemptive, it is likely that these stochastic planners are preemptively disabling a transmitter that is about leak, leading to enhanced leakage mitigation.

On the other hand, while the planners reduce the total number of observed leakage points in comparison to the naive approach, the amount of leakage points detected is still relatively high. One explanation is that these trials encompass some adversarial cases in which a transmitter attempts to straddle the boundary of the RDZ, i.e.,\ the transmitter continues to oscillate between leaking and compliant behavior. Since \Tool{} does not record the history of a transmitter, the periodic behavior circumvents \Tool{}'s basic leakage mitigation policy. To alleviate these adversarial cases, one could incorporate a strike policy into their RDZ, i.e., after $x$ amount of strikes, the offending transmitter may be penalized (e.g.,\ disabled for a period of time).

Nonetheless, based on the results, the HTN-based planners outperform both the PPO-based policy and the naive approach, while offering greater transparency than their RL counterparts as we can easily trace the planning path through the HTN. We observe improved performance among the stochastic HTN methods for certain trials, indicating that stochastic behavior can potentially enhance our planning policies. Furthermore, we observe a significant reduction in leakage points (nearly $10\%$ in Trial 1) and leaked signal strength (approximately $100$ dBm in Trial 1) across trials. Note that the signal results in Figure~\ref{fig:leakage} are presented in log scale (base $10$). Most importantly, we emphasize that \Tool{}'s oversight greatly reduces amount of leakage in comparison to non-agent based environments.

\begin{figure}[tbp!]
	\centerline{\includegraphics[width=0.8\columnwidth]{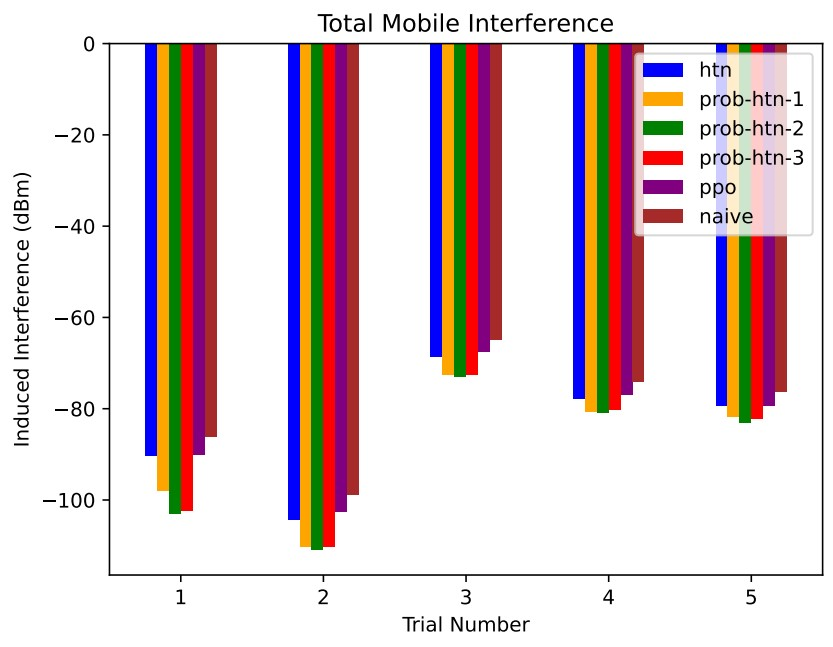}}
	\caption{Observed Mobile Interference. A lower value indicates reduced interference, while a higher value indicates increased interference.}
	\vspace{-0.7cm}
	\label{fig:interference}
\end{figure}

\subsection{Mobile Interference}
We set the mobile interference threshold to $-70$ dBm for this evaluation. Therefore, if the mobile interference exceeded $-70$ dBm, this constituted an RDZ violation. While weak, signals at $-70$ dBm are likely to impact nearby transmitters, users, and test experiments. We collected and recorded the aggregate mobile interference observed and the SINR for the mobile transmitter. In this evaluation, we aimed to measure \Tool{}'s ability to preserve communication (i.e., maintaining adequate SINR and reducing mobile interference). The procedure and setup were nearly identical to the previous evaluation; however, two POWDER partitions ($18 $ nodes total) were deployed and split among $3600$ and $3610$ MHz respectively.

\begin{figure}[tbp!]	\centerline{\includegraphics[width=0.8\columnwidth]{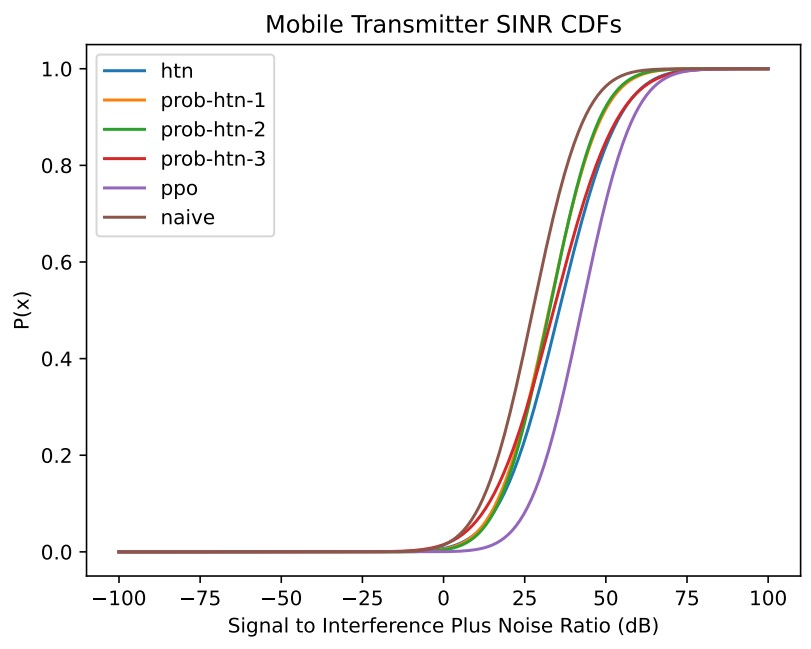}}
	\caption{Observed Mobile SINR. Larger SINR values correspond to improved communication performance, while smaller values indicate degraded communication capabilities.}
	\label{fig:cdf}
 	\vspace{-0.6cm}
\end{figure}

The results of the evaluation are summarized in Figure~\ref{fig:interference} and Figure~\ref{fig:cdf}. We omit the results from the random policy as the transmitter uptime was significantly lower than the other planning approaches (see Table~\ref{tab:uptime}). Based on the results in Figure~\ref{fig:interference}, the stochastic HTN planning methods significantly outperformed all other planning methods; additionally, all planning methods consistently outperformed the naive approach. The consistent decrease in the interference is apparent by the smaller (more negative) interference values obtained by the various planning methods in comparison to the naive approach. Like in the previous evaluation, the stochastic HTN methods' improved interference policy appears to stem from greater state space exploration, with the agent landing in states that have preemptive properties; however, the benefits of the stochastic approaches over their deterministic counterpart are more blatant in this case. Given this, adopting a small perturbation (e.g.,\ $\epsilon = 0.1$) can lead to a noticeable boost in interference reduction and communication preservation. In essence, based on the results, in most trials, \Tool{} outperforms the naive approach with a decrease in total interference of nearly $20$ dBm in some cases.

As for the observed mobile SINR, we observe a small difference between the \Tool{} and the naive case. To highlight the performance differences, we plot the SINR results for the mobile transmitter as a CDF (see Figure~\ref{fig:cdf}). We assume that the SINR values are normally distributed according to the sample mean and the sample standard deviation of the collected results. For all approaches, most of the collected SINR measurements of the mobile transmitter are above $25$ dB, which indicates that the transmitter's transmissions are strong and communication is preserved. Note that PPO outperforms all other planning methods significantly. One reason for this may be due to skewed credit assignment. In this case, the provided reward function may be asymmetrically promoting the mobile SINR above all other metrics; hence, the agent learns a policy that prioritizes the mobile SINR, e.g., always adjusting the mobile transmit frequency to promote the mobile communication channel. Although more evaluations may shed light on this descrepancy, the motivations behind the planning trajectory are opaque, unlike the HTN which is human interpretable. Like the previous evaluations, these results indicate that \Tool{}'s planning-based approaches outperform the naive approach by preserving a higher level of communication for mobile transmitters. We emphasize that \Tool{}'s oversight greatly reduces the magnitude of interference in comparison against non-agent based environments.

\begin{table}[!h]
\begin{tabular}{|c|c|c|c|c|c|c|c|}
\hline
 \ & {\it HTN} & {\it HTN 1} & {\it HTN 2} & {\it HTN 3} & {\it PPO} & {\it Random} & {\it Naive}  \\ \hline
 Uptime & $1.0$ & $1.0$ & $1.0$ & $1.0$ & $1.0$ & $\approx 0.51$ & $1.0$ \\ \hline
\end{tabular}
\caption{Average Mobile Uptime across Trials.}
\label{tab:uptime}
\end{table}



%% file: conclusion.tex


\section{Conclusion}
\label{sec:conclusion}
We present \Tool{}, an autonomous RDZ manager, and validated its design through a proof-of-concept prototype. While promising, further research is necessary. For example, leveraging online samples to update the internal propagation model's parameters would improve \Tool{}'s generalizability. Additionally, real-world benchmarks are crucial, motivating further research into enhancing testbed infrastructure. Despite its limitations, \Tool{} represents an intersection between a growing interest in autonomous control, digital twin modeling, and dynamic spectrum environments. We argue that systems such as \Tool{} will be critical towards realizing an autonomous RDZ amidst the growing sophistication and complexity of wireless environments.

